\documentclass[oldversion]{aa}
\usepackage{txfonts}
\usepackage{natbib}
\usepackage{graphicx}

\bibpunct[, ]{(}{)}{;}{a}{}{,}

\def\llm{{\sc LLmodels}}
\def\atl{{\sc ATLAS9}}
\def\aatl{{\sc ATLAS12}}
\def\aur{$\theta$~Aur}
\def\logg{\log g}
\def\teff{T_{\rm eff}}
\def\tauros{\tau_{\rm Ross}}
\def\kms{km\,s$^{-1}$}
\def\bz{$\langle B_{\rm z} \rangle$}
\def\emf{$e.m.f.$}

\begin{document}

\title{The Lorentz force in atmospheres of CP stars: $\theta$~Aurigae}

\author{D. Shulyak\inst{1} \and G. Valyavin\inst{2,3}
\and O. Kochukhov\inst{4} \and B.-C. Lee\inst{2}
\and G. Galazutdinov\inst{2,3}
\and K.-M. Kim\inst{2} \and Inwoo Han\inst{2}
\and \\ T.~Burlakova\inst{3}
\and V. Tsymbal\inst{5,1} \and D. Lyashko\inst{5,1} }

\offprints{D. Shulyak, \\
\email{denis@jan.astro.univie.ac.at}}

\institute{Institut f\"ur Astronomie, Universit\"at Wien, T\"urkenschanzstra{\ss}e 17, 1180 Wien, Austria \and
Korea Astronomy and Space Science Institute, 61-1, Whaam-Dong, Youseong-Gu, Taejeon, 305-348, Korea \and
Special Astrophysical Observatory, Russian Academy of Sciences, Nizhnii Arkhyz, Karachai Cherkess Republic, 369167, Russia \and
Department of Astronomy and Space Physics, Uppsala University, Box 515, 751 20, Uppsala, Sweden \and
Tavrian National University, Yaltinskaya 4, 95007 Simferopol, Crimea, Ukraine}

\date{Received / Accepted}

\abstract{
Several dynamical processes may induce considerable electric currents in
the
atmospheres of magnetic chemically peculiar (CP) stars. The Lorentz force,
which results from the interaction between the magnetic field and
the induced currents, modifies the atmospheric structure
and induces characteristic rotational variability of the hydrogen Balmer
lines. To study this phenomena we have initiated a systematic
spectroscopic survey of the Balmer lines
variation in magnetic CP stars. In this paper we continue presentation
of results of the program focusing on the high-resolution spectral
observations of A0p star
\aur\, (HD\,40312). We have detected a significant variability of the
H$\alpha$, H$\beta$, and H$\gamma$ spectral lines during full rotation cycle
of the star.
This variability is interpreted in the framework of the model atmosphere
analysis,
which accounts for the Lorentz force effects. Both the inward and outward
directed Lorentz forces are considered
under the assumption of the axisymmetric dipole or dipole+quadrupole
magnetic field configurations.
We demonstrate that only the model with the outward directed Lorentz force
in the dipole+quadrupole
configuration is able to reproduce the observed hydrogen line variation.
These results present new strong evidences for the presence of non-zero
global electric
currents in the atmosphere of an early-type magnetic star.
}

\maketitle

\section{Introduction}
The atmospheres of magnetic chemically peculiar (CP) stars display the
presence of global magnetic
fields ranging in strength from a few hundred G up to several tens of
kG \citep{landstreet2001}.
In the contrast to the complex, localized and unstable magnetic fields of
the cool stars with an external
convective envelope, magnetic fields of CP stars are organized
at a large scale, roughly dipolar \citep{landstreet2001} or low-order
multipolar \citep{bagnulo} geometries,
which are likely stable during significant time intervals. In fact,
these stars provide a unique natural
laboratory for the study of secular evolution of global cosmic magnetic
fields and other
dynamical processes which may take place in the magnetized plasma.
In particular, the slow variation of the field due to decay
changes the pressure-force balance in the atmosphere via the
induced Lorentz force, that makes it possible to detect it observationally
and establish a number of important constraints
on the plausible scenarios of the magnetic field evolution in early-type
stars.

Several physical mechanisms have been suggested as the sources of non-zero
Lorentz force in the atmospheres
of CP stars: the global field distortion and evolution
\citep{stepien,landstreet1987,valyavin}, the global
drift of charge atmospheric particles under the influence of radiative
forces \citep{peterson}, ambipolar
diffusion \citep{leblanc}. Phenomenological atmosphere models with the
Lorentz force included were also
presented by \citet{madeja,madejb} and \citet{carpenter}. In the light of
these discussions it becomes clear that the study of the Lorentz force in
the atmospheres of CP stars is of fundamental importance for understanding
the nature of intrinsic microscopic processes in magnetized atmospheric
plasma.

The aforementioned studies suggested that the magnetic forces may
lead to significant differences between the atmospheric
structures of magnetic and non-magnetic stars. In some cases, the Lorentz
force may noticeably change
the effective gravity and influence formation of the pressure-sensitive
spectral features, especially the hydrogen Balmer lines.
Some of the H$\beta$ photometric data
\citep{madej1984,musielok} can be considered as an evidence
for the presence of non-force-free magnetic fields. Spectroscopy
of the hydrogen lines also points in this direction.
\citet{kroll} found variability with amplitudes more than 1\% in the Balmer
lines in several magnetic stars.
He showed that at least part of this variability can be attributed to the
presence of a non-zero Lorentz force in stellar atmospheres.

Unfortunately, apart from the study by Kroll, who presented low-resolution
spectroscopic observations of the Balmer line variability with rotation
phase in only a few CP stars, there have been no other systematic
spectroscopic surveys of the Balmer line variability. Such a situation is
due to the fact that the observational aspect of the problem is fairly complex,
and comprehensive understanding
can be obtained only with the help of high-precision, high-resolution
spectroscopic observations. Very weak variations in the Balmer line profiles
(for the majority of stars it is 2\% or less) require reconstruction of the
strongly broadened spectral features with an accuracy of $\sim$\,0.1\%.
Until recently, such a precision could not be reached by high-resolution
spectroscopy. However, the situation has been improved significantly with the
development of highly stable fibre-fed spectrographs that made it possible
to carry out these studies at a more advanced instrumental level. Taking this
into account and following the pioneering work by \citet{kroll} we have
initiated a new spectroscopical search of hydrogen line variability \citep{valyavin05}.

In this paper we present the phase-resolved high-resolution observations
of one of the brightest weak-field
($B \approx 1$\,kG) magnetic CP star \aur. We detected significant
variation of the Balmer line profiles and
interpreted it in terms of the non-force-free magnetic field configuration.
We argue that chemically overabundant spots
in the atmosphere of \aur\, can not produce the observed
variability.

In the next section we describe observations and methods of spectral
processing. Variability of the H$\alpha$, H$\beta$,
and H$\gamma$ lines is illustrated in this section.
Sect.~3 introduces the model which we have employed to interpret
the observations. In Sect.~4 we perform calculation of our stellar model
atmosphere with the Lorentz force. Results are presented in Sect.~5 and
summarized in Sect.~6. General discussion is presented in Sect.~7.

\section{Observations}
\aur\, (HD\,40312) is a broad-lined A0p star with a relatively weak
($\approx$1\,kG) dipolar magnetic field \citep[see][]{wade}. During rotation
the star shows equatorial (phases 0.25, 0.75) as well as polar regions
(phases 0.0, 0.5) of its magnetosphere. As follows from the previous
studies \citep[e.g.][]{valyavin}, the maximum atmospheric perturbation by
the Lorentz force is expected to be observed at the equatorial plane
and is nearly zero at the polar regions. This makes it possible to estimate
the magnetic force term by analyzing the differences between the Balmer line
profiles obtained at different rotation phases.

The observations were carried out with the BOES echelle spectrograph installed
at the 1.8\,m telescope of the Korean Astronomy and Space Science Institute.
The spectrograph and observational procedures are described by \citet{kim}.
The instrument is a moderate-beam, fibre-fed high-resolution spectrograph
which incorporates 3 STU Polymicro fibres of 300, 200, and 80 $\mu$m core
diameter (corresponding spectral resolutions are
$\lambda/\Delta\lambda$\,=\,30\,000, 45\,000, and 90\,000 respectively).
The medium resolution mode was employed in the present study. Working
wavelength range is from 3500~\AA\, to 10\,000~\AA. High throughput of the
spectrograph at 4100--8000~\AA\, wavelength range and its high stability
make it possible to obtain spectra of the Balmer lines with an accuracy of
about $0.2-0.3$\%.

Twenty spectra of \aur\, were recorded in the course of about 20 observing
nights from January 2004 to April 2005. Typical exposure times of a few
minutes allowed to achieve $S/N\sim150$.
Table~\ref{Tphases} gives an overview of our observations.
Throughout this study we use ephemeris derived by \citet{wade}:
\begin{equation}
JD=2450001.881+E\times3.61860,
\end{equation}
where the reference time corresponds to the negative extremum of the
longitudinal field variation.
%-----------------------------------------------------------------
\begin{table}
\caption{Observations of \aur. The first column gives the number of
observation, the second column lists the Julian date (JD) and the last
one is the rotation phase
calculated with ephemeris by \citet{wade}.}
\centering
\begin{tabular}{ccc}
\hline\hline
No. & JD & Rotation Phase\\
\hline
 1 & 2453015.0238  &   0.681\\
 2 & 2453015.9949  &   0.950\\
 3 & 2453020.0157  &   0.061\\
 4 & 2453038.0886  &   0.056\\
 5 & 2453039.1761  &   0.356\\
 6 & 2453039.9192  &   0.562\\
 7 & 2453040.9137  &   0.837\\
 8 & 2453042.9287  &   0.393\\
 9 & 2453046.0285  &   0.250\\
10 & 2453340.0242  &   0.496\\
11 & 2453341.0364  &   0.775\\
12 & 2453341.3455  &   0.861\\
13 & 2453343.0469  &   0.331\\
14 & 2453354.0668  &   0.377\\
15 & 2453354.3565  &   0.456\\
16 & 2453356.0911  &   0.936\\
17 & 2453356.3045  &   0.995\\
18 & 2453457.9726  &   0.091\\
19 & 2453458.0466  &   0.111\\
20 & 2453458.0834  &   0.121\\
\hline
\end{tabular}
\label{Tphases}
\end{table}
%-----------------------------------------------------------------
The spectral reduction was carried out using the image processing program
DECH \citep{gala} as well as MIDAS packages. The general steps are standard
and include cosmic ray hits removal, electronic bias and scatter light
subtraction, extraction of the spectral orders, division by the flat-field
spectrum, normalization to the continuum
and wavelength calibration.

%---------------------------------------------------------------------------------
\begin{figure}
\resizebox{\hsize}{!}{\includegraphics{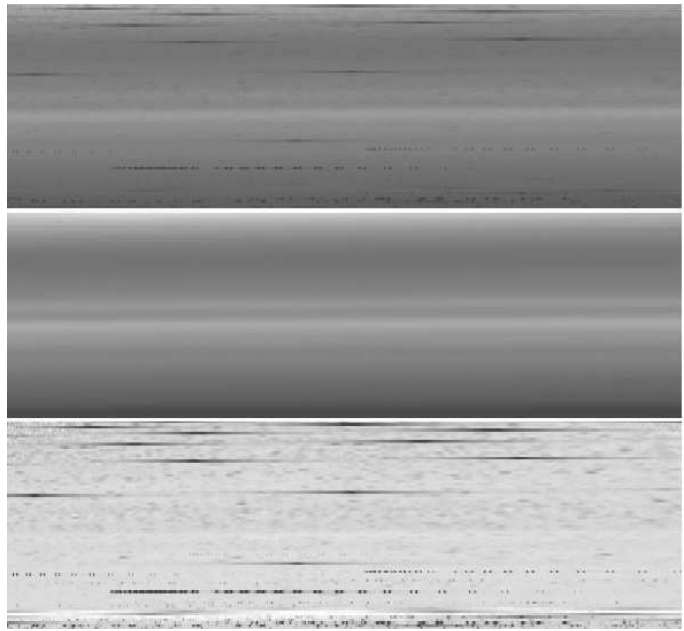}}
%\resizebox{\hsize}{!}{\includegraphics{0000f1.ps}}
%\includegraphics[width=8.8cm,height=6.6cm]
\caption{Examples (from top to bottom) of the initial image of echelle orders,
its 2-D continuum and normalized image.}
\label{Fcont2d}
\end{figure}
%---------------------------------------------------------------------------------

%---------------------------------------------------------------------------------
\begin{figure}
\resizebox{\hsize}{!}{\includegraphics{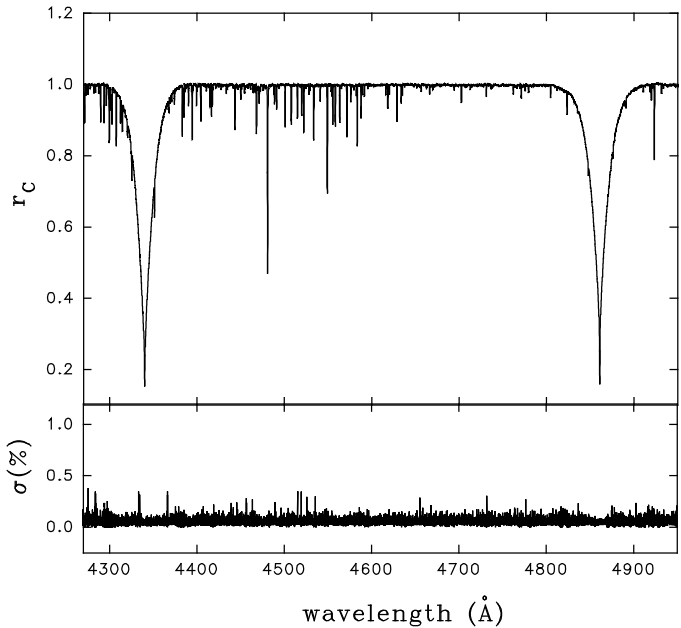}}
\caption{Mean spectrum (the upper plot) and standard deviation
$\sigma$ at the H$\beta$ and H$\gamma$ profile of Vega spectrum.
}
\label{Vega}
\end{figure}
%---------------------------------------------------------------------------------

Normalization to the continuum deserves some additional comments.
In order to obtain uniformly reconstructed continuum in all the spectra
we applied the following technique. After the flat-fielding procedure all
extracted spectral orders of individual
frames were merged into new 2-D (4000x75) images where the echelle orders
(75 orders in total)
where consequently  placed as image rows. Then, with the aid of the median
and Gaussian filters, we
identified and cut out all narrow spectral features in all spectral
orders. Finally, ignoring the Balmer
line regions we fitted all the images using 2-D cubic spline function and
created 2-D continuum images.
Normalized spectra were produced by the division of the initial images by
their corresponding 2-D
continuum images. Examples of the initial image, 2-D continuum derived
from it and normalization results are
presented in Fig.~\ref{Fcont2d}.

Our analysis showed that such a technique makes it possible
to achieve a required high accuracy and stability of the continuum
reconstruction in those places of the observed spectra where the
spectrograph's response function can be described by a monotonic
low-order polynomial 2-D function. For the BOES the most appropriate
region is from 4150 to about 5100~\AA\, that makes
our analysis reliable at the H$\gamma$ and H$\beta$ Balmer lines.
Accuracy of the continuum normalization around these lines
is estimated to be approximately 0.1--0.2\%. In order to illustrate this
conclusion, here we present results of test observations of
Vega which were carried out in different nights in the period from early
March to May of 2006. It is established \citep{peterson2} that
this standard star has the inclination of the rotational axis $i = 4.5 \degr
\pm 0.33 \degr$. Such a small inclination minimizes the probability
to detect any physical variability in the Balmer profiles of Vega's spectrum
allowing to use this star as a standard in our study.
Results of the tests are presented in Fig.~\ref{Vega} where the upper plot
illustrates mean spectrum of the star obtained by averaging the
observed spectra. The lower plot is the dispersion of the result
obtained as standard deviation of the individual spectra from the
mean. As can be seen from the behavior of the dispersion function,
all instabilities of the continuum reconstruction
lie typically below $0.15-0.2$\%.

Also, we would like to note that during these observations
the spectrograph's configuration has been changed many times by service
requirements. Nevertheless, the response function of the spectrograph
showed very good stability despite the fact of these re-configurations.
Analyzing observations of another program
Ap/Bp stars which do not show any significant variability of the Balmer
profiles we also concluded that the response function does not change its
shape for the last three years (2003-2006).

For different reasons, the stability around the other Balmer lines is not
so good compared to the H$\gamma$ and H$\beta$ regions. For example, the H$\alpha$
region is characterized by the presence
of a narrow slope of the response function that
makes it difficult to reconstruct continuum around this line with the
necessary accuracy. However, in this study we decided to include the
H$\alpha$ line also as an illustration.

\begin{figure}
\resizebox{\hsize}{!}{\includegraphics{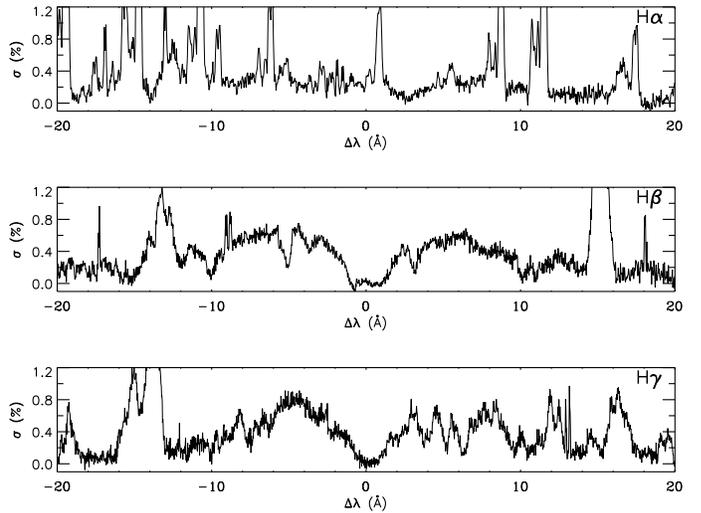}}
%\resizebox{\hsize}{!}{\includegraphics{0000f2.ps}}
%\includegraphics[width=8.8cm,height=6.6cm]{0000f2.ps}
\caption{The standard deviation $\sigma$ at the H$\alpha$, H$\beta$ and H$\gamma$.
The photon noise background of $\sigma\approx0.3$\% is subtracted.}
\label{Fsigma}
\end{figure}

Taking all the above conclusions into account we searched the Balmer line
profile variations using normalized
spectra at the H$\alpha$, H$\beta$, and H$\gamma$
regions. For each of the spectral intervals we examined the standard
deviation $\sigma$ from the mean in the profiles during full rotation cycle
of the star. The standard deviations as function of wavelength are presented
in Fig.~\ref{Fsigma}. Analyzing spectral regions with low lines density
we found that the standard deviation due to the
photon noise and inaccuracies in the spectra processing lie
below the level of 0.3\%. Any deviation above this value indicates the
presence of significant intrinsic variability of Balmer line profiles.

%---------------------------------------------------------------------------------
%---------------------------------------------------------------------------------

The standard deviations at the H$\beta$ and H$\gamma$ reveal the
characteristic
fingerprint first described by \citet{kroll} as the impact of the Lorentz
force: the amplitude rises in the wings and
drops to the line center.
This picture is less clear for the H$\alpha$ line
due to the above reasons and strong distortion of the profile by the
telluric molecular absorptions. (Here we note that
the spectra obtained at different dates on the time base of
about one year are strongly Doppler-shifted relative to each other
due to the Earth's movement. In this connection our study requires rebinding
the spectra to the rest wavelengths and, as a result, all the telluric lines
become randomly shifted in the individual rebined spectra that
strongly complicates analysis of the H$\alpha$ profile).
Nevertheless, the effect is seen in the red wing of
the line (see~Fig.~\ref{Fsigma}).

The narrow features in the standard deviations (H$\beta$ and H$\gamma$ lines,
Fig.~\ref{Fsigma}) come from metal lines
located in wings of the hydrogen line profiles. Strong variability of these lines
indicates the presence of an inhomogeneous surface distribution of
corresponding chemical abundances. This distorting factor
contributes to an additional noise but it can
be isolated and separated from the broad spectral changes in the hydrogen
lines.

\section{Model}
\subsection{General equations and approximations}
In order to model the found variation of the Balmer line profiles
we assume that this variability is caused by the Lorentz force and
follow approaches outlined by \citet{valyavin}.

In the presence of the Lorentz force term the hydrostatic equation reads:
\begin{equation}
\vec{\nabla} P_{\rm total} = \rho \vec{g} + \vec{f}_{\rm L},
\label{Ehydro}
\end{equation}
and
\begin{equation}
\vec{f}_{\rm L} = \frac{1}{c}\vec{j}\times\vec{B},
\label{Efl}
\end{equation}
where $P_{\rm total}=P_{\rm gas}+P_{\rm rad}$ denotes the total pressure,
$P_{\rm gas}$ and $P_{\rm rad}$ are the gas and
radiation field pressures respectively, $\rho$ is the gas density,
$\vec{g}$ is the surface gravity,
$\vec{j}$ and $\vec{B}$ are the surface currents and magnetic field vector
respectively, and $c$ is the speed of light.
The electric currents are determined using the Ohm's law:
\begin{equation}
\vec{j} = \lambda\vec{E}_{\rm \parallel} +
\lambda_{\rm \perp}\vec{E}_{\rm \perp} +
\lambda_{\rm H}\frac{\vec{B}\times\vec{E}_{\rm \perp}}{B},
\label{Ecurrent}
\end{equation}
where $\vec{E}_{\rm \parallel}$ and  $\vec{E}_{\rm \perp}$ are electric
field components directed along and across
magnetic field lines respectively, $\lambda$ is the electric conductivity
in the absence of the magnetic field,
$\lambda_{\rm \perp}$ is the electric conductivity across the magnetic
field lines and $\lambda_{\rm H}$
is the Hall's conductivity.

To simplify solution of the problem, we consider poloidal
surface magnetic field geometry. The non-force-free term giving rise in the
Lorentz force is described via the induced electric field (\emf).
This configuration can be justified in the context of the magnetic field
evolution. For example, distortion of initially force-free configuration
of the magnetic field and respective distribution of the electric field over
the stellar surface may be produced by the global field decay/generation
or by any other, more complicated scenario related, for instance, to
generation of internal toroidal fields by differential
rotation etc. In the present study we do not consider
these details and generalize the problem making no assumption about the
origin of the Lorentz force. Following the approach developed in previous
studies and taking into account the fact that the surface magnetic field of
the majority of CP stars may approximately be described by low-order
axisymmetric poloidal fields, we restrict our model considering only
azimuthal geometry of the induced electric field. In this case
the equations introduced above can be implemented in a 1-D stellar
model atmosphere code.

Finally, the complete set of assumptions used in our modelling can be
summarized as follows:
\begin{enumerate}
\item
The stellar surface magnetic field is axisymmetric and is dominated by
dipolar or dipole+quadrupolar component in all atmospheric layers.
\item
The induced \emf\, has only an azimuthal component, similar to that
described by \citet{wrubel}, who considered decay of the global stellar
magnetic field. In this case the distribution of the surface
electric currents can be expressed by the Legendre polynomials
$P^1_n(\mu)$, where $n = 1$ for dipole,
$n = 2$ for quadrupole, etc.,
and $\mu=\cos\theta$ is the cosine of the co-latitude
angle $\theta$ which is counted in the coordinate system connected
to the symmetry axis of the magnetic field.
\item
The atmospheric layers are assumed to be in static equilibrium and no
horizontal motions are present.
\item
Stellar rotation, Hall's currents, ambipolar diffusion and other dynamical
processes are neglected.
\end{enumerate}

Taking into account these approximations and substituting Eq.~(\ref{Efl})
and Eq.~(\ref{Ecurrent}) into Eq.~(\ref{Ehydro}), we can write the
hydrostatic equation in the following form
\begin{equation}
\frac{\partial P_{\rm total}}{\partial r} =
-\rho g \pm \frac{1}{c} \lambda_{\rm \perp}
\sum_n c_n P^1_n(\mu) \sum_n B^{(n)}_{\rm \theta}.
\label{EhydroR}
\end{equation}
Obtaining this equation we used the superposition principle for field
vectors and the solution of Maxwell equations for each of the multipolar
components following \citet{wrubel}. We also suppose that
$\vec{E}\perp\vec{B}$. Here $c_n$ represents the effective electric field generated by
the $n$-th magnetic field component at the stellar magnetic equator
and $B_{\rm \theta}$ is the horizontal field component.
The signs ``+'' and ``--'' refer
to the outward and inward directed Lorentz forces, respectively.
Equation~(\ref{EhydroR}) can be solved if the plasma conductivity
$\lambda_{\rm\perp}$ is known.

We note that the values of $c_n$ are free parameters to be found
by using our model.
These values represent the fundamental characteristics which
can be used for building self-consistent models of the
global stellar magnetic field geometry and its evolution. Thus, an indirect
measurement of these parameters via the study of the Lorentz force is of
fundamental importance for understanding the stellar magnetism.

\subsection{Calculation of plasma conductivity}
Calculation of the electric conductivity $\lambda_{\rm \perp}$ can be
carried out using the Lorentz collision model where
only binary collisions between particles are allowed. The detailed
description and basic relationships of this approach is given in
\citet{valyavin}.
Improving the latter work, here we calculate the electric conductivity
including all available charged particles.
The precise calculations require direct evaluation of the effective
stopping force acting between
two charged particles of types $k$ (background particle) and $i$
(test particle):
\begin{equation}
F_{\rm eff}^{(ik)} = \frac{4 \pi (Z_i e)^2 (Z_k e)^2 n_k
\ln\Lambda}{M v_{k,\rm T}^2} f_{ik}\left(\frac{v_i}{v_{k,\rm T}}\right).
\end{equation}
Here $n_k$ and $v_{k,\rm T}$ are the concentration and the most probable
thermal velocity of particles
of type $k$, $Z_i$ and $Z_k$ are charges of the particles, $M$ is the reduced
mass and $\ln\Lambda$ is the classical Coulomb logarithm.
The Chandrasekhar function $f(x)$ has the following form
\citep{chandrasekhar}:
\begin{equation}
f(x) = \frac{2}{\sqrt{\pi}}x^{-2}\left[\int_0^x e^{-x^2} dx -
x e^{-x^2}\right].
\end{equation}
In the case of the interaction between charged and neutral particles we
use the elastic collision cross-sections,
\begin{equation}
\sigma_k = \pi a_0^2 \left(\frac{m_k}{m_{\rm p}}\right)^{-2/3},
\end{equation}
where $a_0$ is the Bohr radius and $m_{\rm p}$ is the proton mass.

Generally, at the uppermost
atmospheric layers where the cyclotron frequency of the conducting particles
is much higher than their mean free-path times, effects of magnetoresistivity
on $\lambda_{\rm \perp}$ redirect induced horizontal electric currents
into Hall's currents \citep[see][]{valyavin} which are ignored in our model.
In the deepest layers, the
effects of the magnetic field influence on the conductivity become
negligible and electric currents can be described by ordinary Ohm's law
for the non-magnetic case. In the intermediate atmospheric layers, however,
the situation becomes more complicated. Under the influence of horizontal
magnetic field, the contribution of ionized particles to the
conductivity is strongly stratified by the magnetic field that
produces the {\it bimodal shape} of the Lorentz force term \citep{valyavin}.

Computations of the electric conductivity under the aforementioned assumptions
enables us to present the solution via unknown induced \emf\, ($c_n$) and
magnetic field determined from observations. A preliminary analysis
\citep[see also][]{valyavin} showed that amplitudes of the
pressure-temperature
modification with the Lorentz force depend mainly on the induced \emf\
Magnetic field strength also influences
the amplitudes but mainly affects the
positions of the perturbed bimodal area on the Rosseland
optical depth scale. For the very small magnetic
fields the perturbation of the atmosphere is limited to small Rosseland
optical depths above the region where the wings of the Balmer lines form.
Increasing the magnetic field strength shifts the perturbation
toward the deeper regions, which contribute to the formation of hydrogen
lines. In this process the two maxima of the bimodal atmospheric perturbations
pass one after another through the zone of Balmer line formation, giving
rise to maximum perturbations to the Balmer line formation around
horizontal local magnetic field $B_{\theta}$ of 400--500~G and 10000~G
with a significant gap between $B_{\theta} = 1$~kG and $6$~kG.
In terms of global, nearly dipolar magnetic fields these areas of the
most effective contribution of the horizontal magnetic field to the Balmer
line formation
correspond to approximately 1~kG and 20~kG magnetic stars. Taking into account
that the majority of magnetic stars have magnetic fields weaker than
10~kG we predict that the maximum amplitudes of the Balmer line variation
due to the Lorentz force are expected to be among magnetic Ap/Bp stars
with magnetic fields between 0.4~kG and 1--2~kG.
The \aur\, is one of such stars.

\subsection{Model atmospheres with Lorentz force}
Our calculations were carried out with the stellar model atmosphere
code \llm\, developed by \citet{shulyak}.
The code is based on the modified \atl\, \citep{kurucz13} and \aatl\,
\citep{kuruczA12} subroutines as well as
on the spectrum synthesis package described by \citet{tsymbal}. \llm\,
is written in Fortran\,90 and uses
the following general approximations (which are typical for
many 1-D model atmosphere tools):
\begin{itemize}
\item
the plane-parallel geometry is assumed;
\item
the Local Thermodynamic Equilibrium (LTE) is used to calculate the atomic
level populations for all chemical species;
\item
the stellar atmosphere is assumed to be in a hydrostatic equilibrium;
\item
the radiative equilibrium condition is fulfilled.
\end{itemize}
The code incorporates the so-called line-by-line (LL) method of the
bound-bound opacity calculations \citep{shulyak}.
This technique allows us to account for individual stellar abundance
pattern, which is important in the present study of \aur\,
because this star shows non-solar abundances and an inhomogeneous horizontal distribution of some
chemical elements. In this case accurate treatment of lines opacity in
model atmospheres is needed to ensure correct calculation of model structures with
individual abundance patterns.

The new module was written to compute electric
conductivity and magnetic pressure. Since the effective gravity is a
function of atmospheric depth,
additional changes in the solution of the hydrostatic equation and in the
mass correction routines were made.
The hydrostatic equation is solved and presented in terms of monochromatic
optical depth scale $\tau_{\rm 5000}$ as an independent variable.
At each iteration the code calculates electric conductivity in all
atmospheric layers using all available charged and neutral plasma particles.
The conductivity is then used to evaluate magnetic
contribution to the effective gravity and to execute mass correction
procedure.

As can be seen from Eq.~(\ref{EhydroR}), in the case of the outward
directed Lorentz force, there is some critical value of $c_n$
which may produce unstable solution.
Such models can not be considered in the hydrostatic equilibrium approximation
introduced above and were assumed to be non-physical
in our calculations. Thus, for each set of models, we adopted $c_n$ values
as to ensure static equilibrium.

In addition, the following calculation settings have been used:
the atmosphere is sliced into 72 layers equally
spaced along the Rosseland optical scale height $\tauros$,
from ${\tauros = -6.875}$ to ${\tauros = 2.0}$.
The number of frequency points used for the flux integration procedure was
495\,000 in the 500--50\,000\,\AA\,
spectral region. The initial atomic line list taken from VALD
(Vienna Atomic Line Database) \citep{vald1,vald2} contains information about
21.6 million atomic lines, including lines originating from the predicted
energy levels. This line list was used as an input
for the preselection procedure in the \llm\, code. We adopted the selection
threshold $\ell_{\rm\nu}/\alpha_{\rm\nu} \geqslant 1$\%,
where $\alpha_{\rm\nu}$ and $\ell_{\rm\nu}$ are the continuum and line
absorption coefficients at the given frequency $\rm\nu$.
The preselection enabled us to reduce the total number of lines used for the
line opacity calculations to about 525\,000.

%--------------------------------------------
\begin{figure}[!t]
\resizebox{\hsize}{!}{\includegraphics{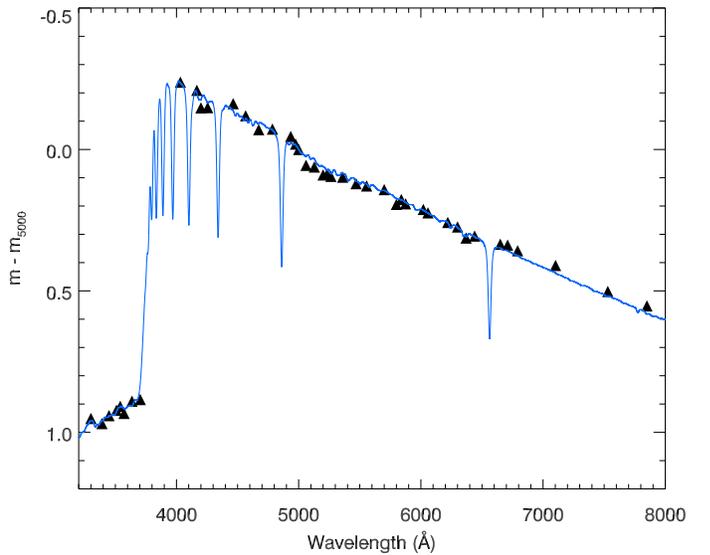}}
\caption{Comparison of the observed (symbols) and computed (solid line)
energy distribution of \aur.
Theoretical model corresponds to $\teff=10\,400$~K, $\logg=3.6$, and has
been convolved with a
$FWHM$\,=\,20\,\AA\, Gaussian filter.}
\label{Fenergy}
\end{figure}
%--------------------------------------------
\begin{figure}[!t]
\resizebox{\hsize}{!}{\includegraphics{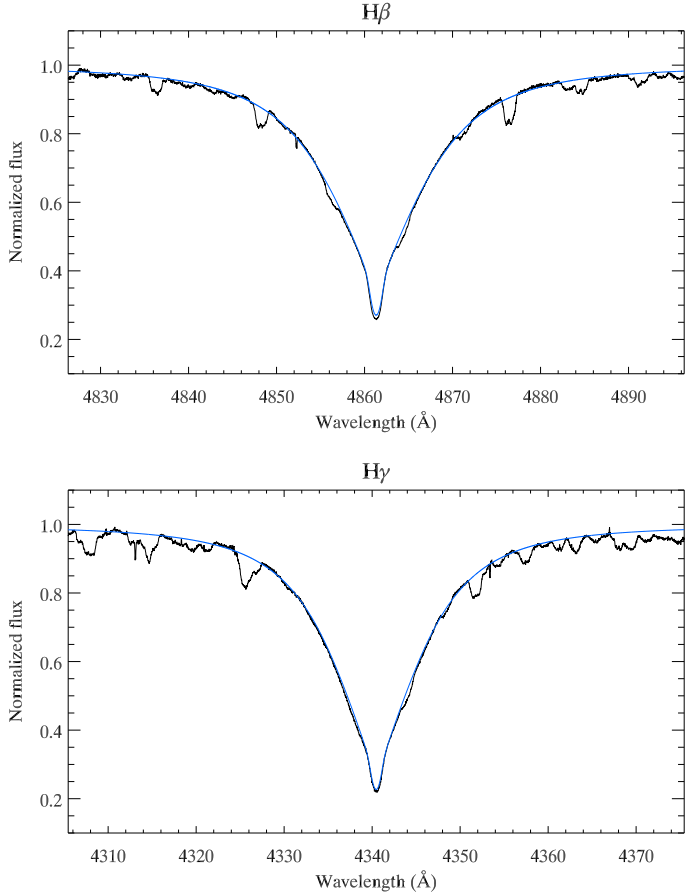}}
\caption{Comparison of the observed and computed H$\beta$
(\textit{upper panel}) and H$\gamma$
(\textit{lower panel}) profiles. Thick line shows observations at
phase 0.995, thin line corresponds to synthetic spectra.}
\label{Fhydlines}
\end{figure}
%--------------------------------------------

\section{Results}
\subsection{Model atmosphere parameters of \aur}
The model atmosphere parameters, $\logg$ and $\teff$, were determined
using spectrophotometric energy distribution \citep{adelman}
and theoretical fit of the H$\beta$ and H$\gamma$ line profiles. For this
purpose we have chosen observations at phase 0.995 (polar region) where the
influence of the Lorentz force is small and the stellar atmosphere is expected to be close to
the standard non-magnetic model structure.
%We note that there may exist some offset between spectroscopic and real
%gravity since the radial component of the Lorentz force is zero only exactly at the magnetic pole of
%the star. But we neglect this offset since in the following we apply
%synthetic spectrum calculations to model the
%relative profile variations only.
The chemical composition as well as the projected rotational velocity
$v\sin i=55$~\kms\ were taken from \citet{kuschnig},
who used multi-element Doppler Imaging technique to derive surface
maps for eight elements in \aur.
Table~\ref{Tabundances} gives abundance patterns for four
representative rotation phases.
%--------------------------------------------
\begin{table}
\caption{The surface-averaged abundances ($\log N/N_{\rm total}$)
of \aur\, for the four rotation phases \citep{kuschnig}
and averaged over the whole stellar surface (last column).}
\centering
\begin{tabular}{cccccccc}
\hline\hline
Element & 0.00  & 0.25  & 0.50  & 0.75  & average\\
\hline
He      & -2.32 & -2.32 & -2.40 & -2.32 & -2.34\\
Mg      & -5.28 & -5.27 & -5.35 & -5.50 & -5.35\\
Si      & -3.35 & -3.27 & -3.09 & -3.22 & -3.23\\
Ti      & -7.52 & -7.61 & -7.67 & -7.61 & -7.60\\
Cr      & -5.14 & -5.35 & -4.99 & -4.75 & -5.06\\
Mn      & -5.50 & -5.64 & -5.49 & -5.34 & -5.49\\
Fe      & -3.86 & -3.83 & -3.63 & -3.69 & -3.75\\
Sr      & -8.43 & -8.43 & -8.36 & -8.19 & -8.35\\
\hline
\end{tabular}
\label{Tabundances}
\end{table}
%--------------------------------------------
Synthetic Balmer line profiles were calculated using the {\sc Synth}
program \citep{piskunov}. The program incorporates recent improvements
in the treatment of the hydrogen line opacity \citep{barklem}.
The stellar energy distribution and Balmer lines are best approximated
with the following parameters: $\teff=10\,400\pm300$~K,
$\logg=3.6\pm0.05$. Note, that such a high accuracy of the determined
parameters is just an internal accuracy obtained from our technique which
we used to fit the data. Real parameters
may be slightly different from the obtained ones due to various systematic error sources, but this does not play
a significant role in our study.
Comparison of the observations and model predictions are presented
in Fig.~\ref{Fenergy} and Fig.~\ref{Fhydlines}. Fundamental parameters
of \aur\, suggest that this star is significantly
evolved from the ZAMS. From the comparison with theoretical evolutionary
tracks \citet{hrmag} find \aur\, to be at the
very end of its main sequence life.

Recent studies by \citet{zeeman_paper1} and
\citet{zeeman_paper2,zeeman_paper3} showed that the effects of Zeeman
splitting and polarized radiative transfer
on model atmosphere structure and shapes of hydrogen line profiles are
less than 0.1\% for magnetic field intensities
around 1~kG and thus can be safely neglected in the present investigation.

\subsection{Magnetic field geometry}
To calculate the Lorentz force effects it is essential to specify the
magnetic field geometry (see Eq.~(\ref{EhydroR})).
The first longitudinal magnetic field measurements were obtained
for \aur\, by \citet{borra} using
H$\beta$ photopolarimetric technique. The authors observed
a smooth single-wave \bz\ variation with rotation phase
and concluded that it is probably caused by a dipole inclined to the
rotation axis of the star. \citet{wade} presented high-precision
longitudinal field measurements of this star. They confirmed and improved
results of \citet{borra}. In particular, deviation of the \bz\ variation
from the purely sinusoidal magnetic curve has been found
\citep[see Fig.~12 in][]{wade}.

Following these works, we have approximated magnetic field topology of
\aur\ by a combination of the dipole and axisymmetric quadrupole magnetic
components. We have also assumed that the symmetry axes of the dipole and
quadrupole magnetic fields are parallel. Thus, the model parameters
include the polar strength of the dipolar component
$B_{\rm d}$, relative contribution of the quadrupole field
$B_{\rm q}/B_{\rm d}$, magnetic obliquity $\beta$,
and inclination angle $i$ of the stellar rotation axis with respect to
the line of sight. The last parameter is best estimated independently,
from the usual oblique rotator relation connecting stellar radius,
rotation period and $v\sin i$. Hipparcos parallax $\pi=18.83\pm0.81$~mas
and $\teff=10\,400\pm300$~K yield $R=5.1\pm0.4\,R_{\odot}$. This
leads to $i=51\degr\pm6\degr$, which is in good agreement with
$i=$\,50\degr--60\degr\ derived in Doppler imaging studies
\citep{rice,kuschnig}.

The remaining free parameters of our model for the magnetic field geometry
were determined with the least-squares fit of the observed \bz\ variation
\citep{borra,wade}. We have examined ``positive'' (quadrupolar field has
two positive and one negative poles)
and ``negative'' (the two negative and one positive poles) quadrupolar field
configurations. We found that the lowest $\chi^2_\nu\equiv\chi^2/\nu$ corresponds to
the negative quadrupolar configuration with
$B_{\rm q}/B_{\rm d}\approx-2$
($\chi^2_\nu=1.14$, $B_{\rm d}=1.4$~kG, $\beta=78\degr$),
while the purely dipole model gives $\chi^2_\nu=2$
($B_{\rm d}=1.3$~kG, $\beta=73\degr$).

\subsection{Effects of horizontal abundance distribution}
\begin{figure}
\resizebox{\hsize}{!}{\includegraphics{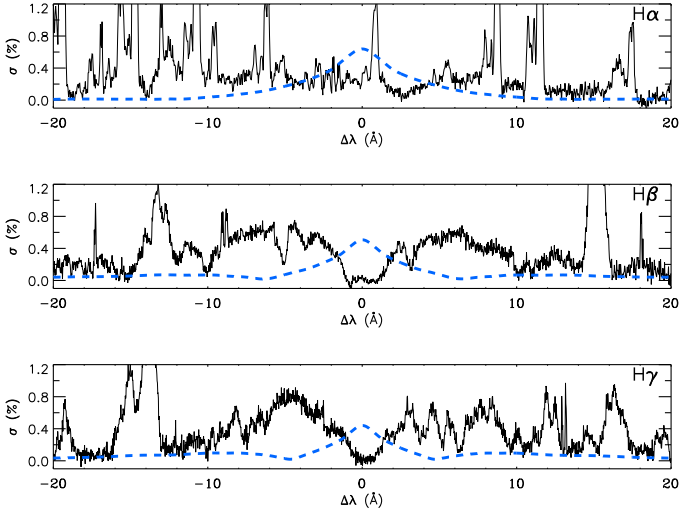}}
%\resizebox{\hsize}{!}{\includegraphics{0000f3.ps}}
%\includegraphics[width=8.8cm,height=6.6cm]{0000f3.ps}
\caption{The standard deviation $\sigma$ at the H$\alpha$,
H$\beta$ and H$\gamma$ lines.
Observations are shown by solid lines. The dashed lines illustrate
variability
due to the horizontal abundance inhomogeneities.}
\label{FsigmaABN}
\end{figure}

In order to distinguish effects of the magnetic pressure from the ones of
the abundance distribution, we calculated model atmospheres for each
of the four representative phases using individual abundances from
Table~\ref{Tabundances}. The synthetic profiles of the
hydrogen lines as well as the standard deviation of the profiles
due to the variable chemical abundances at the listed phases
were then calculated. The results are presented in
Fig.~\ref{FsigmaABN}. As one can see, effects due to a non-uniform abundance distribution
are totally different from the observed one: the strongest
variability occurs in the line core whereas the line wings are not
affected much. This allows us to conclude that the chemical spots do
not lead to the observed Balmer line variations. This fact is in
concordance with the results obtained by \citet{kroll}, who showed
that the found shape of the Balmer line variability can not be
reproduced by metallicity or temperature variations, but can be
considered in the frame of changes in the pressure structure of
the stellar atmosphere. Finally, the average abundances from
Table~\ref{Tabundances} (last column) were used in the rest of
model atmosphere calculations.
%--------------------------------------------
%--------------------------------------------

\subsection{The Lorentz force}
In order to fit the observed variations of the line profiles, both the
inward and outward directed Lorentz forces
were examined through the model atmosphere calculations.
The actual magnetic input parameters of
computations with \llm\, code include the sign of the Lorentz force,
magnetic field modulus $B$, and
the product of two sums $\sum c_n P^1_n \sum B^{(n)}_{\rm\theta}$.
Following the procedure described by \citet{valyavin}, we take the
latter two parameters to be disk-averaged at the individual rotation phases.
We used this simplified procedure because
a direct integration of the model over the stellar surface
is a very time-consuming process under our approaches.
The most general and precise way to produce a disk-integrated model spectrum is to
\begin{itemize}
\item[i] calculate (for given $c_n$ and the polar strength
$B^{(n)}_{\rm p}$) a number of local model atmospheres and synthetic spectra
for all surface areas from the magnetic equator to the pole,
\item[ii] integrate the local spectra over the visible stellar hemisphere
for a set of rotational phases and given orientation of the stellar
rotational and magnetic axis.
\end{itemize}
At the present stage of our studies we are not yet ready to carry out
such computations. Thus, our basic modelling strategy is to replace
the flux from the disk-integrated model which should be computed for each observed
phase with the flux from a single atmosphere model from \llm, with parameters
which represent approximately the disk-averaged means.
%and leave this question for further special consideration.

On the other hand, our additional study revealed that even the strongest
possible intensities of the Lorentz force do not change significantly
the limb darkening in computations with the magnetic model compared to
the non-magnetic one. Therefore, for the first investigation
we used the above simplified approach. This approximate method essentially
assumes linear response of the properties of magnetic atmospheres to changes
in the input parameters.

Finally, the Lorentz force in the cases of dipole and quadrupole magnetic
field geometries can be written as
\begin{equation}
\frac{\partial P_{\rm total}}{\partial r} = -\rho g \pm \frac{1}{c}
\langle \lambda_{\rm \perp} \rangle c_{1}  \left\langle\left(P^1_1(\mu) + \frac{c_{2}}{c_{1}}P^1_2(\mu)\right)
\left(B^{(1)}_{\rm\theta} + B^{(2)}_{\rm\theta}\right)\right\rangle,
\label{EhydroMean}
\end{equation}
where braces mean the disk-averaging operation
($\langle \lambda_{\rm \perp} \rangle$ here means the ``effective''
conductivity calculated with the magnetic field strength averaged over the stellar disc).
Equation (\ref{EhydroMean})
can be written in a simplified form by introducing the effective acceleration
$g_{\rm eff}$ as a sum of gravitational
and magnetic accelerations so that
\begin{equation}
\frac{\partial P_{\rm total}}{\partial r} = -\rho g_{\rm eff}.
\label{EhydroGeff}
\end{equation}

Results of the disk-averaging procedure of the magnetic parameters are
illustrated in Table~\ref{Tav} where we adopted
$B_{\rm q}/B_{\rm d}=0.0$ (dipole) and $-2.0$ (dipole+quadrupole).
Several sets of models with the outward and inward directed Lorentz force
and $B_{\rm q}/B_{\rm d}$ ranging from 0.5 to $-2.0$ with a step of 0.5
were then calculated. The best fit to the shape of the observed hydrogen line variation
was finally found for the following parameters: $c_2/c_1=2.5$ and
$B_{\rm q}/B_{\rm d}=-2.0$.

Numerical tests showed that, in order to
reproduce the amplitudes of the observed standard deviations due to the
profile variations in case of the inward directed Lorentz force, the
effective
electric field should be $c_1=2\times10^{-10}$~CGS units for the dipolar
configuration and $c_1=1\times10^{-10}$~CGS units for the dipole+quadrupole
model. In case of the outward directed Lorentz force these values should be
$c_1=3.5\times10^{-11}$~CGS units and
$c_1=1\times11^{-11}$~CGS units, respectively.
Increasing $c_1$ further leads to unstable solution. This occurs
for both inward and outward directed Lorentz forces. In the latter case this
means that magnetic force directed outwards becomes stronger
than the gravitational force, numerically forcing $P_{\rm gas}<0$. For example,
the critical value of the effective electric field
is $c_1 \approx 2\times10^{-11}$~CGS units for dipole+quadrupole
configuration. In spite of the fact that inward directed Lorentz force increases pressure
of stellar plasma, the enormous increase of $c_1$ causes a failure of some
numerical algorithms implemented in the model atmosphere code (mainly hydrostatic
equation). This also causes numerical problems with the interpolation of partition
functions for some elements. The estimated critical value for inward directed force
is $c_1 > 2\times10^{-10}$~CGS units for dipole+quadrupole configuration.

\begin{table}
\caption{Results of the surface averaging procedure for the dipole and
dipole+quadrupole ($B_{\rm q}/B_{\rm d}=-2.0$, $c_2/c_1=2.5$)
magnetic field configurations. Field strengths are given in Gauss.}
\centering
\begin{tabular}{c|cc|cc}
\hline\hline
\multicolumn{1}{c|}{Rotation} & \multicolumn{2}{c|}{dipole} & \multicolumn{2}{c}{dipole+quadrupole} \\
\cline{2-5}
Phase & $\langle B \rangle$ & $\langle P^1_1 B_{\rm\theta} \rangle$ & $\langle \sum B^{(n)}\rangle$ & $\langle \sum \frac{c_n}{c_1}P^1_n \cdot \sum B^{(n)}_{\rm\theta} \rangle$\\
\hline
0.056 & 856 & 436 & 2158 & -2652 \\
0.061 & 854 & 438 & 2147 & -2636 \\
0.091 & 842 & 449 & 2077 & -2533 \\
0.111 & 833 & 458 & 2016 & -2449 \\
0.121 & 829 & 463 & 1982 & -2406 \\
0.250 & 812 & 480 & 1530 & -1913 \\
0.331 & 863 & 429 & 1332 & -1836 \\
0.356 & 886 & 406 & 1289 & -1878 \\
0.377 & 905 & 388 & 1261 & -1931 \\
0.393 & 920 & 373 & 1240 & -1987 \\
0.456 & 965 & 331 & 1188 & -2198 \\
0.496 & 975 & 321 & 1177 & -2256 \\
0.562 & 955 & 340 & 1199 & -2145 \\
0.681 & 852 & 439 & 1358 & -1827 \\
0.775 & 806 & 485 & 1613 & -1989 \\
0.837 & 813 & 479 & 1833 & -2224 \\
0.861 & 821 & 470 & 1921 & -2329 \\
0.936 & 853 & 439 & 2142 & -2627 \\
0.950 & 858 & 434 & 2169 & -2668 \\
0.995 & 865 & 427 & 2210 & -2734 \\
\hline
\end{tabular}
\label{Tav}
\end{table}

%--------------------------------------------
\begin{figure}
\resizebox{\hsize}{!}{\includegraphics{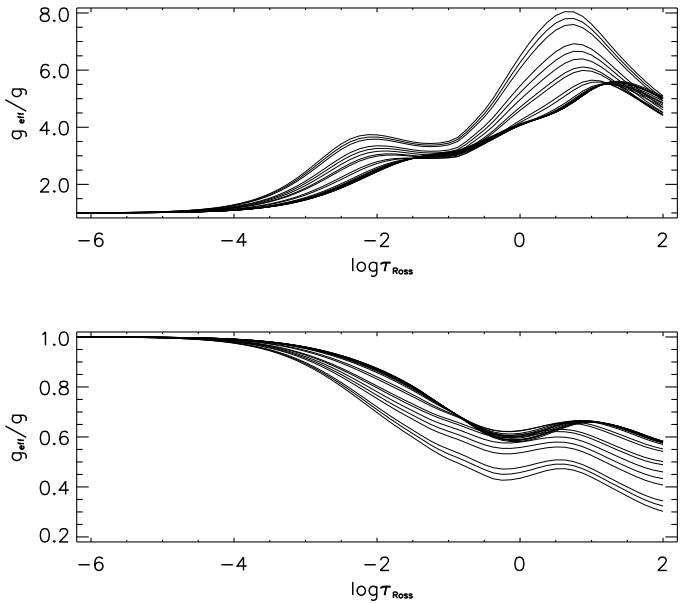}}
\caption{Run of the effective acceleration with Rosseland optical depth
in the atmosphere of \aur\ obtained for all 20 phases of the star's rotation.
The calculations were carried out for inward directed (upper plot)
and outward directed (lower plot) Lorentz force
for dipole+quadrupole magnetic field configuration.}
\label{Fgeff}
\end{figure}
%--------------------------------------------
%\begin{figure}
%\resizebox{\hsize}{!}{\includegraphics{0000f5.eps}}
%%\resizebox{\hsize}{!}{\includegraphics{0000f5.ps}}
%%\includegraphics[width=8.8cm,height=6.6cm]{0000f5.ps}
%\caption{Same as in Fig.~\ref{FsigmaIN} but for the outward directed Lorentz force.}
%\label{FsigmaOUT}
%\end{figure}
%--------------------------------------------

Figure \ref{Fgeff} illustrates resulting dependence of $g_{\rm eff}$
on the Rosseland optical depth in the atmosphere of \aur\ ($B_{\rm q}/B_{\rm d}=-2, c_2/c_1=2.5$).
%The maximum perturbation of the
%atmosphere by the magnetic pressure
%locates at the continuum ($\log\tau_{\rm Ross} \approx 0$) and deeper.
The formation region of the central parts of Balmer lines
($\log\tau_{\rm Ross}<-1$ and above)
demonstrates strong changes in the effective gravity during rotation
that results in the Balmer line variations. Examination of the purely dipolar
model gives similar results.

As follows from Eq.~(\ref{EhydroMean}), the resulting disc-averaged
atmospheric perturbations due to the Lorentz force depend
on the competition between the effective electric conductivity
$\langle \lambda_{\rm \perp} \rangle$ and the product
$\langle \sum \frac{c_n}{c_1}P^1_n \cdot \sum B^{(n)}_{\rm\theta} \rangle$
averaged over the disc. The electric conductivity
$\langle \lambda_{\rm \perp} \rangle$ across the magnetic field lines
of force is reduced in comparison to the non-magnetic conductivity
by the full disc-averaged magnetic field $\langle \sum  B^{(n)}
\rangle $ (the magnetoresistivity effect, see $\S$~2 in \citet{PI66};
\citet{CO57}; \citet{SC50}). In our dipolar magnetic field model
$\langle B \rangle $ (and the magnetoresistivity as a
result) varies with an amplitude of about 10\% about the average value,
which is almost twice smaller than the variation of
the weighted-average of the horizontal magnetic field (see Table~\ref{Tav}).
Therefore, in case of the dipolar geometry, variation of the Balmer lines
is produced mainly by phase variation of the horizontal magnetic field.
In the dipole+quadrupole configuration however, the variation of
conductivity also plays noticeable role
due to significant rotational modulation of $\langle \sum  B^{(n)}
\rangle $ (Table~\ref{Tav}), which explains the influence of magnetoresistivity
on the Balmer line variations for this geometry.

\subsection{Comparison with the observations}
\begin{figure}
\resizebox{\hsize}{!}{\includegraphics{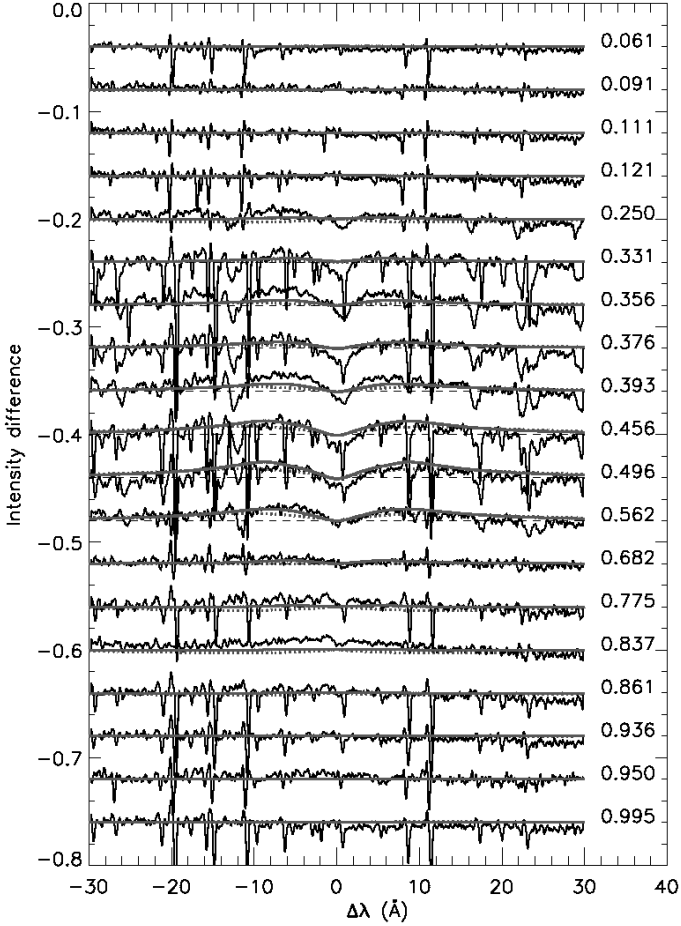}}
%\resizebox{\hsize}{!}{\includegraphics{0000f6.ps}}
%\includegraphics[width=8.8cm]{0000f6.ps}
\caption{Residual profiles of the H$\alpha$ line for each of the observed
phase relative to the phase 0.056. Thin solid lines are observations, dotted lines
are theoretical profiles calculated
for the dipole magnetic field configuration (inward directed Lorentz force),
thick solid lines are profiles
calculated for the dipole+quadrupole magnetic field configuration with
$B_{\rm q}/B_{\rm d}=-2.0$ and $c_2/c_1=2.5$ (outward directed Lorentz force).
The residual spectra for consecutive phases are shifted in the vertical
direction. The thin dashed line gives zero level for each spectrum.
%Observations have been smoothed by 15 points second-order polynomial fit.
}
\label{FHadiff}
\end{figure}

\begin{figure}
\resizebox{\hsize}{!}{\includegraphics{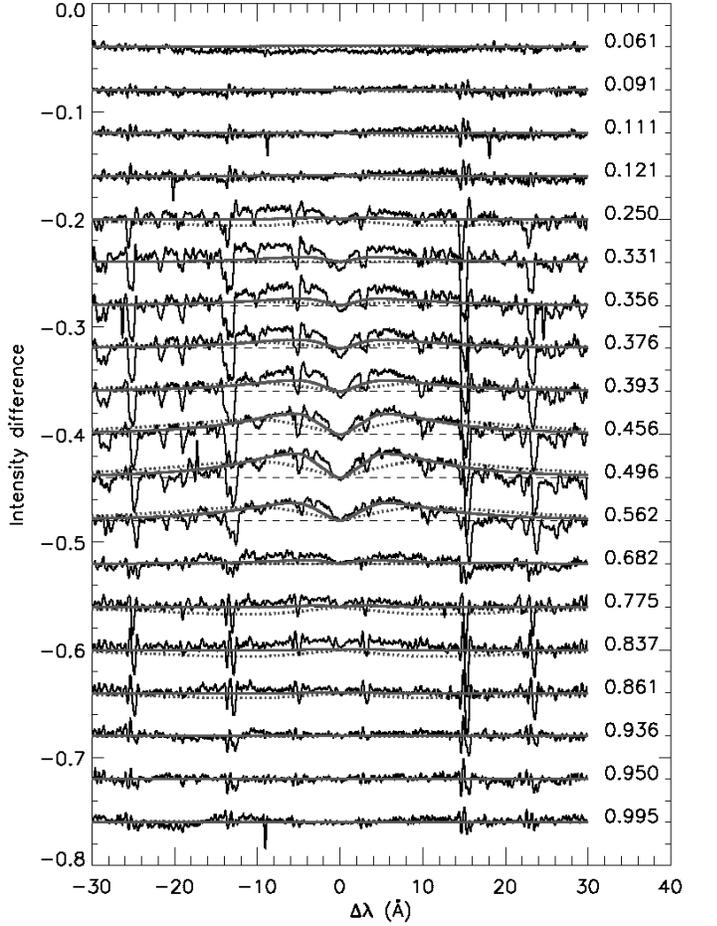}}
\caption{Same as Fig.~\ref{FHadiff} but for H$\beta$ line.}
\label{FHbdiff}
\end{figure}

\begin{figure}
\resizebox{\hsize}{!}{\includegraphics{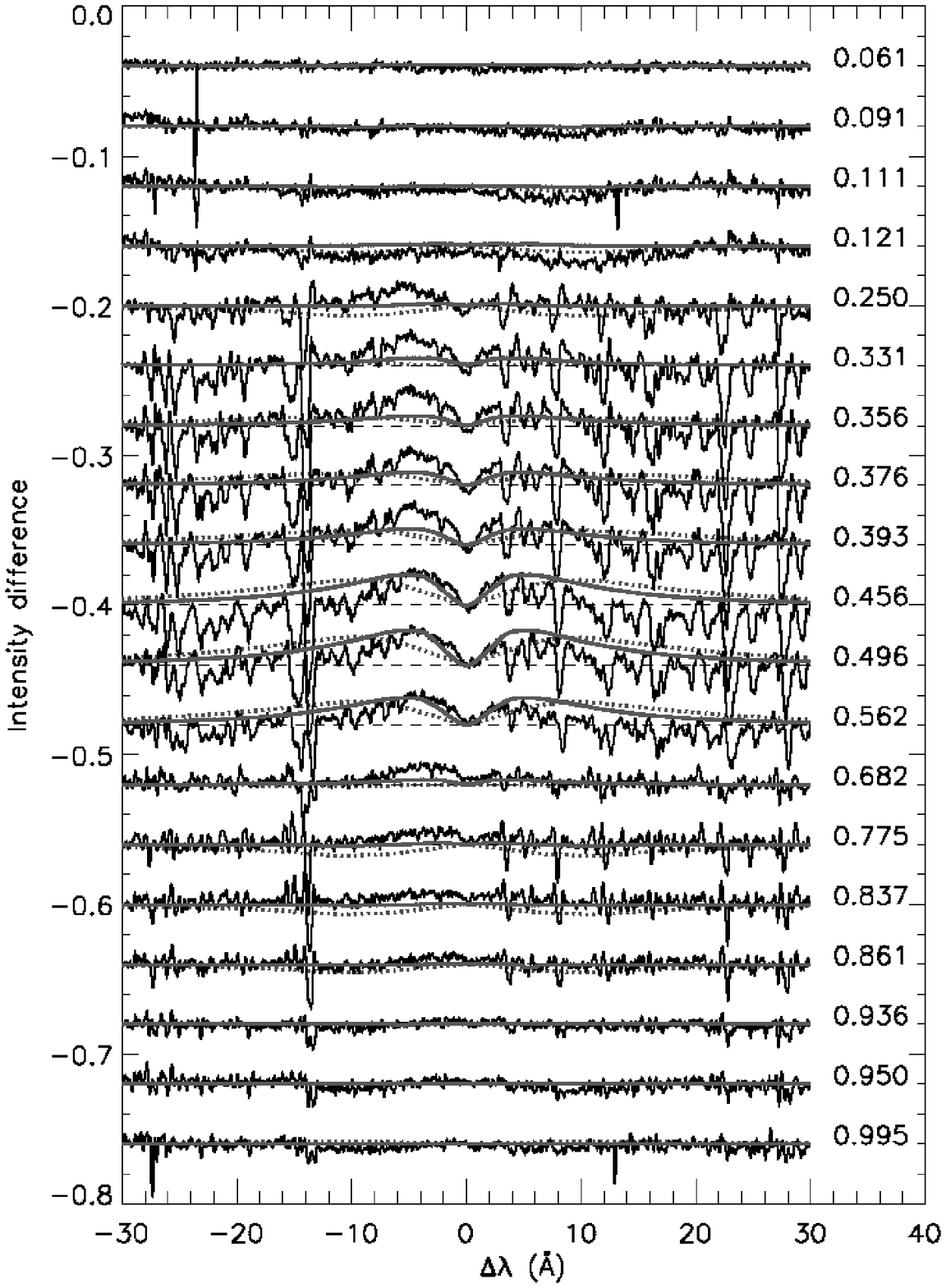}}
\caption{Same as Fig.~\ref{FHadiff} but for H$\gamma$ line.}
\label{FHgdiff}
\end{figure}

We compared observations and theoretical predictions for each
of the rotation phases. To minimize effects of systematic errors in
the modelling procedure, we have examined residual theoretical and observed
Balmer lines, which are obtained by subtracting
a spectrum at a reference phase ($\phi = 0.056$ where the Balmer profiles
have the largest widths) from all the other spectra.
Figures~\ref{FHadiff}, \ref{FHbdiff} and \ref{FHgdiff}
illustrate residual H$\alpha$, H$\beta$ and H$\gamma$ line profiles
for each of the observed rotation phases. The
positive sign of the residuals implies that the lines at the current phase are narrower
than those obtained at the reference phase. Here we present
only those calculations, which are able to reproduce the observed positive
deviations with satisfactory accuracy.
This is achieved for the purely
dipolar geometry of the magnetic field with the inward directed Lorentz
force (dotted lines in the figures) and by examination of the dipole+quadrupole geometry
with the outward directed force (thick solid lines).

The most noticeable effect is seen at phases between $\phi~=~0.250$
and $\phi = 0.6$. However, both models fit the data only in the phase region between
$\phi = 0.45$ and $\phi = 0.6$. We attribute this disagreement
to our assumption of the axisymmetric
magnetic field geometry (in reality the field can be decentered or
distorted). Despite this discrepancy we suppose that our
model reproduces general phase dependence found in observations.

Comparing results of the fit obtained from both models we conclude that
observations are better reproduced in the case of the outward directed
Lorentz force for the dipole+quadrupole magnetic field geometry.
As can be seen from Fig.~\ref{FHbdiff}, \ref{FHgdiff}
the inward directed force gives too wide line wings and is unable to
describe the line cores of the observed residual spectra.
In contrast, the outward directed Lorentz force shows
better agreement with observations. Such a difference between the two
models
results from the fact that at the upper atmospheric layers of the Balmer
line formation regions
the inward directed Lorentz force changes the pressure-temperature balance
more effectively than the outward directed one (see Fig.~\ref{Fgeff}).
Increasing $g_{\rm eff}$, the inward directed Lorentz force makes the Balmer
line profiles non-realistically wider than the observed ones.
This makes the outward directed magnetic force more preferable in our
analysis.

The best fit obtained with $B_{\rm q}/B_{\rm d}=-2.0$
and $c_2/c_1=2.5$ agrees with the longitudinal field modelling results. The
outcome of the H$\gamma$ line profile analysis (Fig.~\ref{FHgdiff}) is in
satisfactory agreement with the H$\beta$ line modelling. In the case of the
H$\alpha$ line (Fig.~\ref{FHadiff}) it is difficult
to distinguish predictions of the two models.

%Besides, our model assumptions may not be
%quite realistic because of more complicated nature of the Lorentz force
%observed (see the discussion below). Having no possibility to consider
%it in this investigation we leave this question for further considerations.
%We should only note that despite some definite weakness in our
%assumptions (the ideal symmetry of the field geometries and so on)
%we got new very strong observational evidences for the presence of
%the non force-free magnetic fields in atmospheres of CP stars that makes it
%possible to discuss them in more detail.
%--------------------------------------------
%--------------------------------------------
%--------------------------------------------
%--------------------------------------------

Now, before general discussion, we would like to clarify some points in
the above modelling approaches. The observed changes in Stark profiles of the
Balmer lines demonstrate monotonic single-wave variation during full
rotation cycle of \aur. Originally, however, we expected
another picture of the variations due to predicted dominant dipolar
component of the magnetic field in this star.
According to previous studies \citep[and references therein]{valyavin}
we expected that if a magnetic star with centered dipolar field shows the observer both
positive and negative parts of its magnetosphere during rotation (that takes
place in \aur), the hydrogen lines should demonstrate double-wave
variations during full rotation cycle. To resolve this difficulty, we
have included the quadrupole component, which is still not well-established
for \aur, but is plausible for the reasons mentioned
above. In the contrast to the purely dipolar magnetic field,
the combination of dipole+quadrupole enables us to derive
a satisfactory fit to the observed behavior of the Balmer line
variations. As follows from Table~\ref{Tav},
we are able to obtain a configuration of the dipole+quadrupole magnetic field
for which the maximum visible horizontal component
is located at those phase points where the varying longitudinal magnetic field
exhibits extremum (not at the crossover).
Such a configuration gives the single-wave variation of the Lorentz force
and reproduces our observations.

Another important point is related to different
orientations of the Lorentz force obtained for the two examined geometries of
the magnetic field. For
both geometries with the opposite-oriented Lorentz force
the sign of the resulting Balmer line residual variations is unchanged at the extremal
point of the phase curve ($\phi \approx 0.5$). In this connection
we note again (see above), that the Stark widths variations in
the purely dipolar
model are mainly produced by the horizontal magnetic field that
provides the minimum inward directed Lorentz force at this phase due to
the minimum of visible horizontal component (see Table~\ref{Tav}). The Stark widths of
the lines are reduced at this phase in comparison to the other phases.
In contrast, the
dipole+quadrupole geometry of the magnetic field provides
significant horizontal component at this point and minimum
magnetoresistivity that results in the maximum outward directed magnetic force
term. As a result, the Stark widths of the Balmer lines are also reduced
at this phase relatively to all the other phases of the stellar rotation
even implementing the opposite orientation of the Lorentz force.

Despite the fact that a fairly good agreement with observations is obtained, we do not
claim that our geometric fit is absolutely correct.
For example, our assumption about essentially
poloidal centered magnetic field may be wrong. Intuitively we may assume,
that decentered dipole may also produce the Balmer line profile
variations which can be similar to the observed ones. Besides,
we do not exclude the presence of more complicated dynamical,
non-evolutionary
processes such as meridional circulation which can also produce similar
observable features.
The problem is too complicated, and the answer about the true model parameters
(also about the true nature of the Lorentz force) can be obtained only when
the exact geometry of the surface field in \aur\ will be
established. In this study we only demonstrate that the observations
{\it can be
described with simple geometrical approaches about
magnetic field and induced electric currents}. We therefore conclude that
the main achievement of this paper is not our explanation of the
behavior of the phase-resolved Balmer line profiles,
but our argumentations for the presence of the detectable Lorentz force
in the atmosphere of \aur.

%It is likely that the
%mismatch is due to the Hall's current effects which are neglected in our
%calculations. As it is shown by \citet{valyavin}, the Hall's currents become important
%in all atmospheric layers for the magnetic field strength greater than
%$\sim$1~kG and for the stellar parameters close to those of \aur. The resulting
%picture of the distribution of the electric currents acquires a complex 3-D structure
%and can not be treated in the present 1-D models.

\section{Conclusions}
With the aim to probe the presence of the Lorentz force in atmospheres of magnetic CP stars, we have acquired
new high-precision spectral observations for one of the bright magnetic stars, \aur. The hydrogen
line profile variations are investigated using high-resolution spectra covering the full
rotation cycle of the star. We have detected and studied variability
of the Stark-broadened profiles of H$\alpha$, H$\beta$, and H$\gamma$.
Several evidences for the presence of significant Lorentz force in atmosphere
of \aur\ are found:
\begin{itemize}
\item
The characteristic shape of the variation during full rotation cycle of the star corresponds to those
described by \citet{kroll} and other authors \citep[and references therein]{valyavin} as a result of an impact of a substantial Lorentz force.
\item
Numerical calculations of the model atmospheres with individual abundances
demonstrate that the surface chemical spots can not produce the observed
variability of the hydrogen line profiles of \aur.
\item
Our model shows good agreement with the observations
if the outward directed magnetic force is applied assuming the
dipole+quadrupole magnetic field configuration with the
induced effective equatorial electric field of $c_1=1\times10^{-11}$~CGS units.
\item
The dipole+quadrupole magnetic model with the quadrupolar strength twice
stronger than the dipolar one reproduces behavior of the phase-resolved H$\beta$ and H$\gamma$ spectra.
\end{itemize}

\section{Discussion}
In the above considerations we did not examine any mechanisms of the
magnetic force generation. We have restricted ourselves to the purely
phenomenological and geometrical description of the problem. Let us finally
discuss physical processes which could create the Lorentz force in the
atmospheres of magnetic CP stars.

\subsection{Secular evolution of the global magnetic field}
It is well known that any variation of the global stellar magnetic field
(related, for instance, to a global field decay) leads to the development of
the induced electric currents in conductive atmospheric layers. The Lorentz
force, which appears as a result of the interaction between the magnetic
field and electric currents, may affect the atmospheric structure and
influence the formation of spectral lines causing their rotational variability.
The sign of the observed Lorentz force (outward or inward directed)
is also found to be important. For example, following \citet{wrubel}, who was
the first to present numerical calculations of the decay of
poloidal interior magnetic field, the decay-induced electric
field $E_{\rm\phi}$ have essentially azimuthal form:
\begin{equation}
 E_{\rm\phi} \sim g(r) P^1_n(\mu),
\end{equation}
where $g(r)$ is a scalar radial function of a distance $r$ from stellar center ($r = [0,R_\star]$).
Considering only the dipole component, the induced
currents do not change sign at the stellar surface,
achieving maximum strength at the magnetic equator (where the magnetic field
lines of force is horizontal) and vanishing at the magnetic poles ($\theta = 0\degr, 180\degr$).
In this case the Lorentz force
%, which can also be presented through the radial function
%\begin{equation}
% f_l \sim -g (\frac{d g}{d r} + \frac{g}{r})
%\end{equation}
%\citep[see realtionships for the necessary components]{wrubel},
corresponds
to the case of the outward directed decay-induced force term at any
non-polar area of the stellar surface. Therefore, determination of the sign of the Lorentz force
provides an important constraint which is able to restrict {\it direction of the field evolution}.

According to \citet{landstreet1987}, who also considered a decay
of an essentially dipolar fossil magnetic field, we should not expect any
signatures
of the Lorenz force in atmospheres of CP stars. In his model magnetic field
has dipolar geometry
throughout a magnetic star and electric field induced at the equator can
be approximated with
the following expression:
\begin{equation}
E_{\rm eq} \sim \frac{R_{\rm \star} B_{\rm eq}}{c t},
\label{Einduce}
\end{equation}
where ${R_{\rm\star}}$ is the stellar radius, $B_{\rm eq}$ is the strength of
the dipole-like surface field at the magnetic equator and $t$ is
the characteristic decay time of the magnetic field. For a typical A0
magnetic star with the surface field of 10~kG this leads to the value
$E_{\rm eq} \sim 10^{-13}$ CGS, which is about 2--3 orders of
magnitude smaller compared to the strength of the electric currents tested
in our model calculations and inferred from the variations of
hydrogen lines. The former theoretical estimate of
$E_{\rm eq}$ is based on the $10^{10}$\,yr characteristic decay time of
nearly dipolar fields. Thus, if the observed strong electric currents
in the atmosphere of \aur\, have the evolutionary nature, we may conclude
that the general concept of a slow decay of the fossil, essentially dipolar,
fields requires for revision. One should examine more realistic structures
of the internal fields, and this may lead to faster field evolution.
As it was recently shown by \citet{mhd}, dynamical stability of the global stellar magnetic
fields is ensured if the interior field configuration includes both poloidal
and toroidal components. Diffusive evolution of such a twisted field structure
may lead to relatively rapid change of the surface field intensity. This,
in turn, induces a noticeable atmospheric Lorentz force during certain
periods in the star's life. In this context advanced evolutionary stage
of \aur\ is remarkable.

Thus, large amplitude of the induced electric currents may indicate
that the poloidal magnetic geometry, dominating at the surface, becomes significantly
distorted inside a magnetic star. Such a distortion is very likely to be related
to dynamical processes of the field evolution and, possibly, interaction
with the core convective zone or differential rotation.

\subsection{Non-evolutionary surface effects}
Generally, any global magnetic topology gives a large collection of
possibilities of generation of the surface electric currents even without invoking
the global magnetic field evolution. There are mechanisms which may produce
significant atmospheric currents even in constant magnetic fields.
For instance, \citet{peterson} considered interaction of the horizontal
component of the magnetic field with a flow of charged particles, drifting
in the atmosphere under the  influence of the radiation pressure. The result
of such an interaction is that drifting particles acquire some horizontal
velocity component. This leads to appearance of the Lorentz force which may
be significant for hot stars ($T_{\rm eff} > 18\,000-20\,000$\,K).
More recently \citet{leblanc} studied a similar physical situation in the
context of ambipolar diffusion. Furthermore, interaction of the magnetic field
and stellar rotation may induce dynamical processes
leading to the development of additional plasma flows (such as meridional
circulations) which are able to create electric currents and the Lorentz force.

We have obtained our results under the assumption that the equatorial
induced electric field is constant in vertical direction. This should be
close to reality if we deal with evolving global magnetic field, whose
variation is small throughout the observable atmospheric layers.
This assumption, however, may
not be applicable if the Lorentz force is related to, for example,
the ambipolar diffusion. Nevertheless,
our results provide an observational test to distinguish these effects.

As it was shown by \citet{leblanc}, the ambipolar diffusion is small for
the weak field stars and increases in stars with the surface magnetic
fields of about 10~kG and higher. It should be noted, however, that we have
considered a relatively weak-field star ($B$\,$\approx$\,1~kG).
Moreover, the {\it inward directed} Lorentz force predicted by the ambipolar
diffusion theory is not supported by our observations, as we have found a
much better
agreement for the {\it outward directed} magnetic force. Nevertheless,
we do not claim that our approach is the only way to address the
problem of the generation of the Lorentz force in the atmospheres of CP stars.
Before making final conclusion about the nature of the Lorentz
force found in \aur, alternative models should be applied to interpret our observational
findings.

\begin{acknowledgements}
We thank J. Landstreet and G. Wade for useful discussions and their interest in this investigation.
We also thankful to J. Braithwaite for useful comments.
This work was supported by INTAS grant 03-55-652 to DS and Austrian Fonds zur Foerderung
der wissenschaftolichen Forschung (P17890).
GV and GG are grateful to the Korean MOST (Ministry of Science and
Technology, grant M1-022-00-0005) and KOFST (Korean Federation of Science
and Technology Societies) for providing them an opportunity to work at KAO through Brain Pool program.
GV acknowledge the Russian Foundation for Basic Research for financial support
(RFBR grant N 01-02-16808).
\end{acknowledgements}


\begin{thebibliography}{100}
\bibitem[Adelman et al., 1987]{adelman}Adelman, S. J., Pyper, D. M., Shore, S. N., White, R. E., \& Warren, W. H. 1989, \aaps, 81, 221
\bibitem[Barklem et al., 2000]{barklem}Barklem, P. S., Piskunov, N., \& O'Mara, B. J. 2000, \aap, 363, 1091
\bibitem[Bagnulo et al., 2002]{bagnulo}Bagnulo, S., Landi Degl'Innocenti, M., Landolfi, M., \& Mathys, G. 2002, \aap, 394, 1023
\bibitem[Borra \& Landstreet, 1980]{borra} Borra E. F., \& Landstreet J. D. 1980, \apjs, 42, 421
\bibitem[Braithwaite \& Spruit, 2004]{mhd}Braithwaite, J., \& Spruit, H. C. 2004, Nature, 431, 819
\bibitem[Carpenter, 1985]{carpenter}Carpenter, K. G. 1985, \apj, 289, 660
\bibitem[Chandrasekhar, 1942]{chandrasekhar}Chandrasekhar, S. 1942, Principles of Stellar Dynamics, University of Chicago Press.
\bibitem[Cowling, 1957]{CO57}Cowling, T. G. 1945, MNRAS, 105, 166
\bibitem[Galazutdinov, 2000]{gala}Galazutdinov, G.A. 1992, Prep. Spets. Astrophys. Obs., 92
%\bibitem[Glagolevskij \& Chountonov, 2001]{glagolevskij}Glagolevskij, Yu. V., \& Chountonov, G. A. 2001, Bull. Spec. Astrophys. Obs., 51, 88
%\bibitem[Hubrig et al., 2000]{hubrig}Hubrig, S., North, P., \& Mathys, G. 2000, \apj, 539, 352
\bibitem[Khan \& Shulyak, 2006a]{zeeman_paper2}Khan, S., \& Shulyak, D. 2006a, \aap, 448, 1153
\bibitem[Khan \& Shulyak, 2006b]{zeeman_paper3}Khan, S., \& Shulyak, D. 2006b, \aap, 454, 933
\bibitem[Kim et al., 2000]{kim}Kim, K. M., Jang, J. G., Chun, M. Y., et al. 2000, Publication of the Korean Astronomical Society, 15S, 119 (in Korean)
\bibitem[Kochukhov et al., 2005]{zeeman_paper1}Kochukhov, O., Khan, S., \& Shulyak, D. 2005, \aap, 433, 671
\bibitem[Kochukhov \& Bagnulo, 2006]{hrmag}Kochukhov, O., \& Bagnulo, S. 2006, \aap, 450, 763
\bibitem[Kroll, 1989]{kroll}Kroll, R. 1989, \rmxaa 2, 194
\bibitem[Kurucz, 1993a]{kurucz13}Kurucz, R. L. 1993, Kurucz CD-ROM 13, Cambridge, SAO
\bibitem[Kurucz, 1993b]{kuruczA12}Kurucz, R. L. 1993, in Proc. IAU Coll. 138, Peculiar versus Normal Phenomena in A-type and Related Stars, eds. M. Dworetsky, F. Castelli, \& R. Faraggiana, ASP Conf. Ser., 44, 87
%\bibitem[Kurucz, 1970]{kuruczSAO}Kurucz, R. L. 1970, SAO Special Report, No. 308
\bibitem[Kupka et al., 1999]{vald2}Kupka, F., Piskunov, N., Ryabchikova, T. A., Stempels, H. C., \& Weiss, W. W. 1999, \aaps, 138, 119
\bibitem[Kuschnig, 1998]{kuschnig}Kuschnig, R. 1998, Ph.D. Thesis, University of Vienna
\bibitem[Landstreet, 1987]{landstreet1987}Landstreet, J. D. 1987, \mnras, 225, 437
\bibitem[Landstreet, 2001]{landstreet2001}Landstreet, J. D. 2001, in Magnetic Fields Across Hertzsprung-Russell Diagram, eds. G.~Mathys, S.K.~Solanki and D.T.~Wickramasinghe, ASP Conf. Ser., 248, 277
\bibitem[LeBlanc et al., 1994]{leblanc}LeBlanc, F., Michaud, G., \& Babel, J. 1994, \apj, 431, 388
\bibitem[Madej, 1983a]{madeja}Madej, J. 1983a, \actaa, 33, 1
\bibitem[Madej, 1983b]{madejb}Madej, J. 1983b, \actaa, 33, 253
\bibitem[Madej et al., 1984]{madej1984}Madej, J., Jahn, K., \& St\c{e}pie\'{n}, K. 1984, \actaa, 34, 419
%\bibitem[Moss, 1984]{moss}Moss, D. 1984, \mnras, 207, 107
\bibitem[Musielok \& Madej, 1988]{musielok}Musielok, B., \& Madej, J. 1988, \aap, 202, 143
\bibitem[Peterson \& Theys, 1981]{peterson}Peterson, D. M., \& Theys, J. C. 1981, \apj, 244, 947
%\bibitem[Peterson et al., 2006]{peterson2}{\bf Peterson, D. M., Hummel, C. A., Pauls, T. A., Armstrong, J. T., Benson, J. A., Gilbreath, G. C., Hindsley, R. B., Hutter, D. J., Johnston, K. J., Mozurkewich, D., Schmitt, H. R. 2006, \nat, 440, 896}
\bibitem[Peterson et al., 2006]{peterson2}Peterson, D. M., Hummel, C. A., Pauls, T. A., et al. 2006, \nat, 440, 896
\bibitem[Pikelner, 1966]{PI66}Pikelner, S. B. 1966, {\it Principles of cosmic electrodynamics}, (in Russian), Moscow: Nauka
\bibitem[Piskunov, 1992]{piskunov}Piskunov, N. 1992, in Stellar Magnetism, ed. Yu. V. Glagolevskij, I. I. Romanyuk (St. Petersburg: Nauka), 92
\bibitem[Piskunov et al., 1995]{vald1}Piskunov, N. E., Kupka, F., Ryabchikova, T. A., Weiss, W. W., \& Jeffery, C. S. 1995, \aaps, 112, 525
\bibitem[Rice \& Wehlau, 1991]{rice}Rice, J. B., \& Wehlau, W. H. 1991, \aap, 246, 195
\bibitem[Schluter, 1950]{SC50}Schluter, A. 1950, Zs.f. Naturforsch, 5a, 72
\bibitem[Shulyak et al., 2004]{shulyak}Shulyak, D., Tsymbal, V., Ryabchikova, T., St\"utz\, Ch., \& Weiss, W. W. 2004, \aap, 428, 993
\bibitem[St\c{e}pie\'{n}, 1978]{stepien}St\c{e}pie\'{n}, K. 1978, \aap, 70, 509
\bibitem[Tsymbal, 1996]{tsymbal}Tsymbal, V. V. 1996, in Model Atmospheres and Spectral Synthesis, eds. S. J. Adelman, F. Kupka \& W. W. Weiss, ASP Conf. Ser., vol. 108, 198
\bibitem[Valyavin et al., 2004]{valyavin}Valyavin, G., Kochukhov, O., \& Piskunov, N. 2004, \aap, 420, 993
\bibitem[Valyavin et al., 2005]{valyavin05}Valyavin, G., Kochukhov, O., Shulyak, D., Lee, B.-C., Galazutdinov, G., Kim. K.-M., \& Han, I. 2005, JKAS, 38, 283
\bibitem[Wade et al., 2000]{wade} Wade, G. A., Donati, J.-F., Landstreet, J. D., \& Shorlin, S.\,L.\,S. 2000, \mnras, 313, 851
\bibitem[Wrubel, 1952]{wrubel} Wrubel, M. H. 1952, \apj, 116, 291
\end{thebibliography}
\end{document}